\documentclass[aps,prl,twocolumn,preprintnumbers,superscriptaddress,10pt,longbibliography]{revtex4-2}

\usepackage[T1]{fontenc}
\usepackage[utf8]{inputenc}

\usepackage{amssymb}
\usepackage{amsmath}
\usepackage{amsfonts}
\usepackage{graphicx}
\usepackage{subfigure}
\usepackage{color}
\usepackage{dcolumn}
\usepackage{hyperref}

\newcommand{\eq}{\begin{equation}}
\newcommand{\eqx}{\end{equation}}
\newcommand{\eqn}{\begin{eqnarray}}
\newcommand{\eqnx}{\end{eqnarray}}
\newcommand{\be}{\begin{equation}}
\newcommand{\ee}{\end{equation}}
\newcommand{\bea}{\begin{eqnarray}}
\newcommand{\eea}{\end{eqnarray}}

\newcommand{\f}[2]{\frac{#1}{#2}}

\newcommand{\LL}{{\mathcal L}}

\newcommand{\eps}{\varepsilon}

\newcommand{\Sg}{\Sigma}

\newcommand{\gm}{\gamma}
\newcommand{\Gm}{\Gamma}

\newcommand{\qq}{\quad\quad}

\begin{document}

\title{Phase transitions in an expanding medium -- hot remnants}

\author{Romuald A. Janik}
\email{romuald.janik@gmail.com}
\affiliation{Institute of Theoretical Physics and Mark Kac Center for Complex Systems Research, Jagiellonian University, Łojasiewicza 11, 30-348  Kraków, Poland}
\author{Matti J\"arvinen}
\email{matti.jarvinen@apctp.org}
\affiliation{Asia Pacific Center for Theoretical Physics, Pohang 37673, Republic of Korea}

\affiliation{Department of Physics, Pohang University of Science and Technology, Pohang 37673, Republic of Korea}

\author{Jacob Sonnenschein}
\email{cobi@tauex.tau.ac.il}
\affiliation{The Raymond and Beverly Sackler School of Physics and Astronomy, Tel~Aviv University, Ramat Aviv 69978, Israel}

\begin{abstract}
We analyze the dynamics of a first order confinement/deconfinement phase transition in an expanding medium using an effective boundary description fitted to the holographic Witten model.
We observe and analyze hot plasma remnants, which do not cool down or nucleate bubbles despite the expansion of the system. The appearance of the hot remnants, the dynamics of their shrinking and subsequent dissolution and further heating up is very robust and persists in such diverse scenarios as boost-invariant expansion with a flat Minkowski metric and cosmological expansion in a Friedmann–Robertson–Walker spacetime.
\end{abstract}

\preprint{APCTP Pre2025 - 004}

\maketitle


\noindent {\bf Introduction.} In this letter we study the behaviour of a physical system which expands and cools, and consequently undergoes a first order phase transition. 
Studying 
an expanding medium is interesting in view of various important applications, in particular an expanding and cooling fireball of quark-gluon plasma in high energy collisions, as well as the expanding of the early universe.
In the present letter we primarily consider boost-invariant expansion providing a simple kinematical implementation of the physics of expanding plasma, but we also discuss uniform cosmological expansion. 

We employ a relatively simple model combining hydrodynamics and a dynamical order parameter or ``mixing fraction'' introduced in~\cite{Janik:2021jbq}. The model was fitted to describe the main features of the holographic Witten model~\cite{Witten:1998zw} (without flavours), that were worked out in~\cite{Aharony:2005bm}, with regard to the physics associated to the phase transitions.
However, the structure of the model is very general and we expect that it can be applied in many other settings. 
Note that the confinement-deconfinement phase transition~\cite{Aharony:2006da,Gursoy:2008bu,Gursoy:2008za,Alho:2015zua,Arefeva:2018hyo,Bea:2021zsu} has been studied  intensively using holography.

Suppose that a physical system has a first order phase transition at $T=T_c$ and starts off in the stable high energy phase (which is the deconfined phase in the Witten model) at $T>T_c$. When the system expands, it cools 
and crosses the phase transition temperature. Then bubbles of the low energy (``confined'') phase start to nucleate. When the temperature falls further below $T_c$, the bubbles expand, coalesce and the even more supercooled deconfined plasma tends to nucleate even more, which is expected to realize the phase transition.

The main result of this letter in part challenges this simple picture: we observe the very generic appearance of hot remnants of the high energy phase, whose temperature remains in the vicinity of $T_c$ despite the expansion of the medium. For this reason, as the plasma is not supercooled, these remnants do not dissipate through bubble nucleation of the low energy phase, but rather shrink. Then the system starts to ``dissolve'' with the remaining decreasing fraction of the deconfined phase being counterintuitively even heated up.
These findings seem very generic and robust and appear in such diverse scenarios as boost-invariant expansion in flat Minkowski space or cosmological expansion in a FRW universe. 
The physics behind this behaviour is an interplay between the hydrodynamics of the high energy phase, latent heat and the dynamics of the order parameter. After introducing the details of our model, we will describe the various stages in the evolution of the remnants, namely the initial ``heating up'', a shrinking stage and subsequent dissolution.


\begin{figure}[t]
	\begin{center}
	\includegraphics[width=\columnwidth]{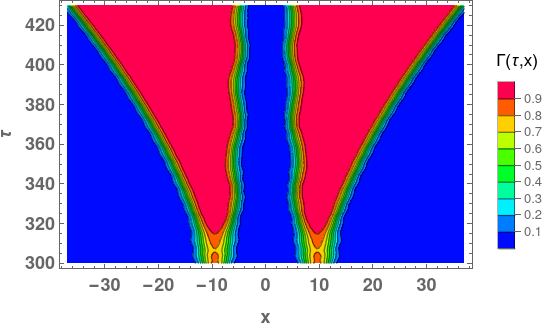}
	\caption{Evolution of the confined ($\Gm=1$) and deconfined phase ($\Gm=0$) from two separated Gaussian perturbations of a boost-invariant expanding deconfined plasma system.}
	\label{fig.twogaussian}
	\end{center}
\end{figure}

\medskip

\begin{figure}[t]
	\begin{center}
	\includegraphics[width=.4\textwidth]{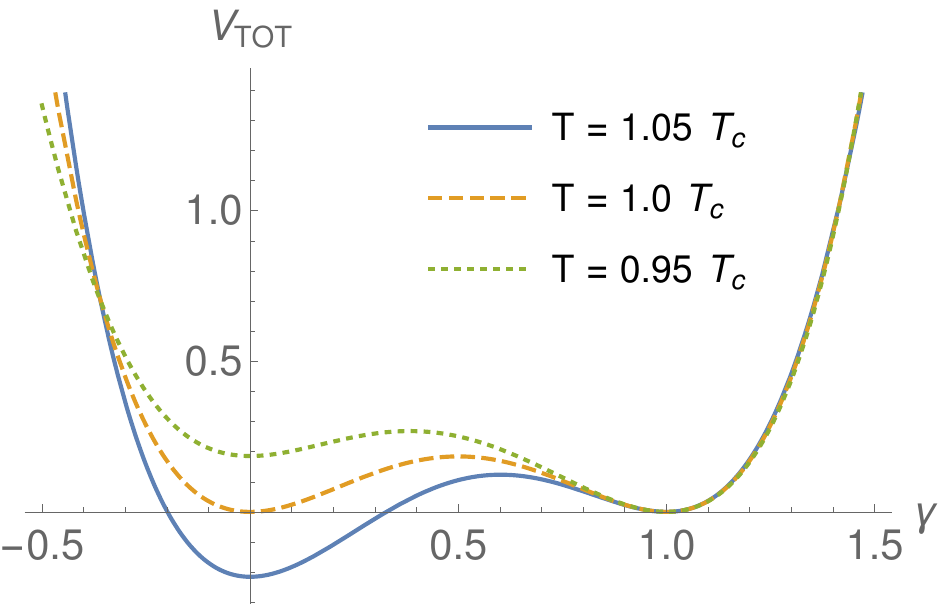}
	\caption{ The potential at $T<T_c$, at $T=T_c$ and $T>T_c$.}
	\label{fig.potential}
	\end{center}
\end{figure}

\noindent {\bf Simplified model of the holographic system.}
As the bulk numerical relativity description of time dependent dynamics in the holographic Witten model is extremely challenging~\cite{Bantilan:2020pay}, in~\cite{Janik:2021jbq}, we introduced a boundary model which 
describes very well the domain wall in the dual field theory, obtained through numerical relativity~\cite{Aharony:2005bm}. The model combines together the hydrodynamic description of the deconfined phase with a simple description  
of the confined phase, by adding an additional dynamical degree of freedom $\gm$ describing the mixing of the two phases (see also~\cite{Bigazzi:2020phm,Bigazzi:2020avc}). Specifically, the total energy-momentum tensor of the theory is a mixture of the hydrodynamic 
tensor of the deconfined phase and the trivial  $T^{\mu\nu}=\eta^{\mu\nu}$ of the confined phase, together with a contribution describing the domain wall ($T^{\mu\nu}_\Sg$):
\eq
\label{e.one}
T^{\mu\nu} = (1-\Gm)\left((\eps+ p)u^\mu u^\nu + p \eta^{\mu\nu}\right) +
\Gm \eta^{\mu\nu} + T^{\mu\nu}_\Sg
\eqx
where $\Gm(\gm)=\gm^2 (3-2\gm)$ is the mixing fraction (see \textit{Supplemental material} for full formulas and rationale for this choice of $\Gm(\gm)$). Here we use units where $p(T)=T^4$, $\eps(T)=3\,T^4$ and the phase transition occurs at $T_c=1$. 
The dynamics of $\gm$ follows from the Lagrangian
\eq
\label{e.two}
\LL = \f{c}{2} (\partial \gm)^2 + V_\mathrm{TOT}(\gm, T)
\eqx
where
\eq
\label{e.three}
V_\mathrm{TOT}(\gm, T) = -(1-\Gm) p(T) -\Gm + \f{c}{2} q_*^2 T^\beta(\gm) \gm^2 (1-\gm)^2 +1
\eqx
Here $c$ is related to the surface tension of the domain wall at $T=T_c$ and $q_*$ sets its width.
The potential for $T<T_c$, for $T=T_c$ and for $T>T_c$ is shown in Fig.~\ref{fig.potential}.
In the case of the Witten model with three-dimensional boundary theory $\beta(\gm) = 2(1+ \Gm)$, while for the four-dimensional (``conformal'' version) $\beta(\gm)=4$~\cite{Janik:2021jbq}.
Note that the above form of the double well potential was fitted only in the neighbourhood of $T_c=1$ up to its first $T$-derivative. For definiteness, we use the above form also away from $T_c$.
The potential $V_\mathrm{TOT}(\gm, T)$ has the interpretation of the Landau free energy of a system with an order parameter~$\gm$. 

One can view the model (\ref{e.one})-(\ref{e.three}) as an extension of the (perfect fluid) hydrodynamics of the deconfined phase (at $\Gm=\gm=0$) by a dynamical order parameter field $\gm$ describing the passage to the confined phase (at $\Gm=\gm=1$). 
The explicit $T$ dependence in the Lagrangian~(\ref{e.two}) induces a very specific coupling between $\gm$ and hydrodynamics 
through the formalism of~\cite{Haehl_2015}. 
This coupling is necessary in order to well describe~\cite{Janik:2021jbq} the domain wall in the Witten model~\cite{Aharony:2005bm}.
Note that we do not have any dissipation terms which are usually added (see e.g.~\cite{Ignatius:1993qn,Hindmarsh:2015qta}) in order to generate friction to determine bubble wall velocity, as our model by itself reproduces correct bubble wall velocities in holographic theories~\cite{Janik:2022wsx}.

\begin{figure}[t]
	\begin{center}
	\includegraphics[width=\columnwidth]{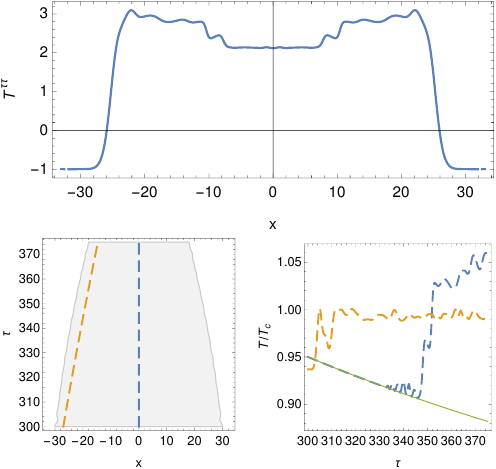}
	\caption{The central plasma region of Fig.~\ref{fig.twogaussian}. Hydrodynamic waves at $T\sim T_c$ at the edges of the central region at $\tau=335$ as seen through the energy density $T^{\tau\tau}$ (top). Evolution of temperature on the wave and at the center (bottom).}
	\label{fig.energytemp}
	\end{center}
\end{figure}

\medskip

\noindent {\bf Boost invariant expansion.}
Boost invariant expansion, first introduced by Bjorken~\cite{Bjorken:1982qr}, can be seen as a way of modelling extremely energetic heavy-ion collisions, where the system undergoes expansion along the longitudinal (collision) axis. 
Consequently, this leads to decreasing energy density and hence also decreasing temperature.
The inherent time-dependence and simplicity led boost-invariant expansion to be a fruitful arena for investigating various dynamical phenomena in holography (see e.g.~\cite{Janik:2005zt,Chesler:2009cy,Heller:2011ju,Heller:2013fn,Gursoy:2015nza,AragonesFontbote:2024lor}).
The expansion automatically drives the cooling system through any phase transitions or crossover to the low energy confined phase.
Technically, one writes the standard Minkowski metric in adapted coordinates
\eq
ds^2 = -d\tau^2 +\tau^2 dy^2 + dx^2
\eqx
with $y$ being the (pseudo-)rapidity along the longitudinal axis and $\tau$ is the proper time in the longitudinal plane. The assumption of boost-invariance amounts to the independence of the dynamics on $y$. 

\medskip

\noindent {\bf Stages in the evolution of the remnant.}
As initial conditions for our evolution we take constant temperature $T=0.95$ and we perturb the $\Gm=\gm=0$ plasma phase by two Gaussians. These mimic two seeds which might arise from thermal or quantum fluctuations which are not encompassed by our dynamical model.
In Fig.~\ref{fig.twogaussian}, we show the subsequent evolution, with increasing domains of the stable low energy phase appearing ($\Gm=\gm=1$). In between these domains, we have a trapped region of supercooled plasma. The boundaries of this region are domain walls which move towards each other. As demonstrated in \cite{Janik:2022wsx}, the moving domain walls are preceded by hydrodynamic waves at $T\sim T_c$. The decrease of the temperature to the supercooled value in the middle (which also decreases with $\tau$) occurs on the wave fronts. The energy which results in heating up of the plasma comes from the latent heat released by the phase transition. The profiles of the energy density and the temporal dependence of the temperature in the middle (blue dashed curves) and on the waves (orange dashed curves) are shown in Fig.~\ref{fig.energytemp}.
Once the waves reach the center, the whole plasma region stops being supercooled.
The duration of this process can be estimated as $\tau_1 -\tau_0 \sim L_0/(2c_s)$, where $c_s$ is the speed of sound and $L_0$ is the initial width. The approximate size of the region after this time is $L_1 \sim L_0 -2 \overline{v_d} (\tau_1 -\tau_0)$, where $\overline{v_d}$ is the average domain wall velocity which can be estimated as in~\cite{Janik:2022wsx} (see also~\cite{Li:2023xto,Wang:2023lam}).

\begin{figure}[t]
\begin{center}
 \includegraphics[width=.4\textwidth]{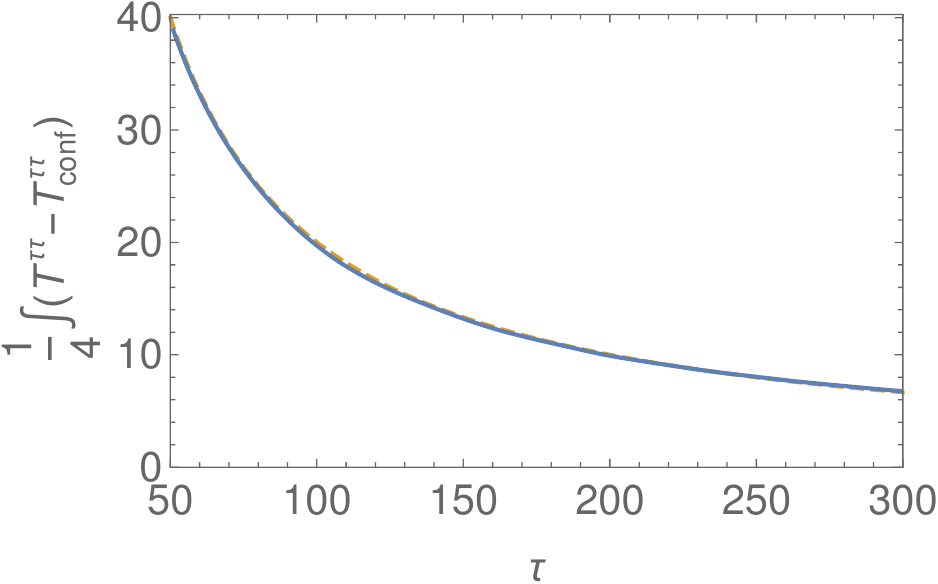}
\caption{The evolution of the size of the shrinking remnant estimated through the integral of $T^{\tau\tau}$. The blue curve is numerical data and the dashed orange curve shows the expected $1/\tau$ decay.}
\label{fig.size}
\end{center}
\end{figure}

After some oscillations, the subsequent evolution of the plasma remnant becomes quite simple. We observe that the temperature is approximately $T_c$ and stays \emph{constant}, while the fluid velocity is quite small. We should understand how to reconcile this constant temperature with the boost-invariant expansion (along the longitudinal coordinate $y$ (not shown)) which normally leads to $T \sim 1/\tau^{\f{1}{3}}$. 
To this end, we can make a rough estimate by considering the conservation of energy with the shrinking of the transverse size of the blob releasing latent heat which can keep the temperature constant. 

In view of a similar computation in the FRW spacetime, it is convenient to use an integrated version of the conservation equation in a general time-dependent geometry~\cite{Clough:2021qlv} 
\eq
\partial_t \left( \int_\Sg d^{D-1}x \sqrt{\tilde{\gm}}\, T^{00} \right) + \f{1}{2} \int_\Sg d^{D-1}x \sqrt{\tilde{\gm}}\, T^{ij} \partial_t \tilde{\gm}_{ij} = 0
\eqx
where $\tilde{\gm}_{ij}$ is the induced metric on the spatial hypersurface $\Sg$.
Neglecting at $0^{th}$ order any $x$ dependence within the blob, we obtain
\eq
\label{e.integrated}
\partial_\tau \left( \tau E \right) + P = 0
\eqx
with
\eqn
\label{e.Ttautau}
E &=& 3p(\tau) L(\tau) - (R - L(\tau))\\
\label{e.Tyy}
P &=&  p(\tau) L(\tau) +  (R - L(\tau))
\eqnx
coming from the components $T^{\tau\tau}$ and $T^{yy}$ integrated over~$x$ and denoting by $R$ the overall transverse size of the system and by $L(\tau)$ the size of the remnant.

If the remnant is in equilibrium with the confined (i.e. low-energy) phase, then $p(\tau)=1$ and we see that the size of the remnant scales like 
\eq
L(\tau) \sim \f{1}{\tau}
\eqx
in agreement with the numerical simulation (see Fig.~\ref{fig.size}).
Incidentally this behaviour at a constant temperature gives an alternative way of satisfying the constancy of entropy per unit rapidity giving
\eq
S \sim \tau L(\tau) T^3 \propto constant \qq \text{as} \qq L(\tau) \sim \f{1}{\tau}
\eqx
An analogous estimate for Bjorken flow~\cite{Bjorken:1982qr} (i.e. standard expansion of plasma $p(\tau)\sim 1/\tau^{\f{4}{3}}$, obtained by setting $L(\tau)=R\equiv L_0$ in equations (\ref{e.Ttautau})-(\ref{e.Tyy})) realizes the constant entropy differently
\eq
S \sim \tau L_0 T(\tau)^3 \propto constant \qq \text{as} \qq T(\tau) \sim 1/\tau^{\f{1}{3}}
\eqx

The above rough estimate of the size of the blob assumed a thin wall approximation which means that all sizes are much larger than the width of the domain wall (set in our model by $q_*$). 
Therefore we can estimate the duration of the shrinking stage as $\tau_2/\tau_1 = L_1/L_{dw}$, where $L_{dw}\sim 1/q_*$ is the typical width of the domain wall.
Once the width of the remnant shrinks to this size, its evolution enters a qualitatively new stage which we now describe.

\begin{figure}[t]
	\begin{center}
	\includegraphics[width=.45\textwidth]{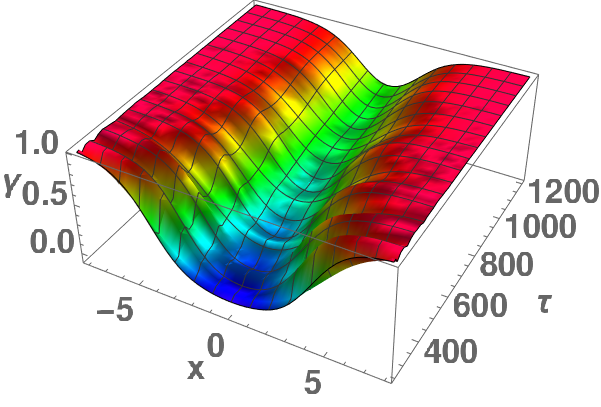}
	\caption{Evolution of $\gm$ when the remnant starts dissolving.}
	\label{fig.gamma_dissolve}
	\end{center}
\end{figure}

\begin{figure}[t]
	\begin{center}
	\includegraphics[width=\columnwidth]{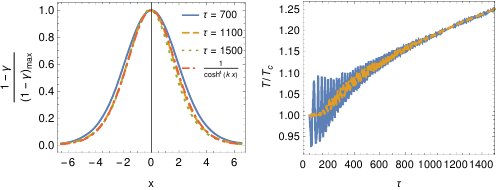}
	\caption{The profile of $1-\gm$ normalized to the same height at different times during the dissolving process, together with the approximate profile (\ref{e.profile}) (left). A slow rise of the temperature (blue)  
 of the dissolving remnant high temperature phase compared with virial theorem prediction (orange) (right).}
	\label{fig.vanishing}
	\end{center}
\end{figure}

The value of $\gm$ now starts to rise towards $\gm=1$, meaning that the fraction of the deconfined high energy phase is decreasing (see Fig.~\ref{fig.gamma_dissolve}). Interestingly enough, the profile of $1-\gm$ seems to preserve the same approximate shape (see Fig.~\ref{fig.vanishing}(left))
\eq
\label{e.profile}
1-\gm(\tau) \sim \f{b(\tau)}{\cosh \left( d(\tau) q_* \f{x}{4} \right)^4}
\eqx
with $b(\tau)$ behaving roughly like $1/\sqrt{\tau}$ and $d(\tau)\sim 1$ in the range of the numerical simulation.
Simultaneously, we observe that the temperature of the residual deconfined plasma slowly \emph{increases} (see Fig.~\ref{fig.vanishing}(right)).
We can understand both of these features in the following way. 
As the expansion is slow enough, the temporal derivatives in the $\gm$ equation of motion are much smaller than the spatial derivatives. Therefore the profile of $\gm$ at each instant of time is essentially determined by the virial approximation
\eq
\label{e.virial}
c\, {\gm'}^2 = 2 V_\mathrm{TOT}(\gm, T)
\eqx
Consequently, the maximum of $1-\gm$ has to occur at a zero of the potential $V_\mathrm{TOT}(\gm_*, T)$. As for the dissolving remnant $\gm \to 1$, this necessarily implies that the temperature must move away from $T=T_c \equiv 1$ and \emph{increase} towards $T>T_c$ (see e.g. the blue curve in Fig.~\ref{fig.potential} which has a zero at $\gm_* \sim 0.35$). 
Indeed, we observe an increase in the temperature in the center of the dissolving blob in accordance to the equation $V_\mathrm{TOT}(\gm(t), T(t))=0$ (see Fig.~\ref{fig.vanishing}(right)). It is important to note, that $T$ in our model denotes the (hydrodynamic) temperature of the deconfined phase, as the confined phase in the Witten model is temperature independent.

For the description of the profile (Fig.~\ref{fig.vanishing}(left)), the relevant part of the potential is the range between $\gm=1$ and the first zero $\gm=\gm_*$.
The approximate shape of the profile (\ref{e.profile}) follows from the virial equation (\ref{e.virial}) with a simple approximation for the potential in that range 
\eq
V_\mathrm{TOT}(\gm=1-g) \sim d^2 q_*^2 g^2 (1 - \sqrt{g/b})
\eqx
In the above way we can understand the structure of the temporal snapshots of the dissolving remnant. In order to get a rough estimate of the temporal dependence, we turn to the entropy current $j^\mu = -\partial_T V_\mathrm{TOT}(\gm, T) u^\mu$ which is conserved in our model as we do not include dissipation (see \textit{Supplemental material}). As in the dissolving remnant configuration the flow velocity is $u^\mu=(1,0,0)$, the entropy conservation reduces to
$
\partial_\tau j^\tau + \f{1}{\tau} j^\tau = 0 
$
which gives
$
s = j^\tau = \f{f(x)}{\tau}
$
Using the formula for the entropy current and expanding $s=-\partial_T V_\mathrm{TOT}$ around $\gm=1$ gives $1-\Gm \sim 1/\tau$ or equivalently $1-\gm \sim 1/\sqrt{\tau}$.


\begin{figure}[t]
	\begin{center}
	\includegraphics[height=5.45cm]{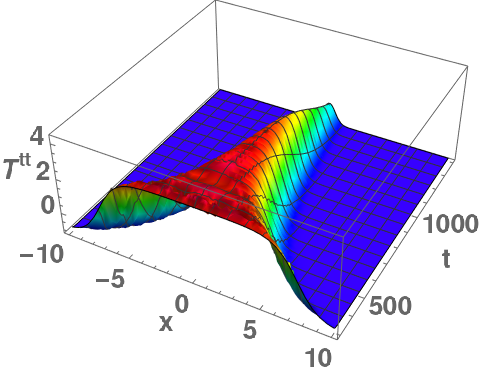}
	\caption{Evolution of $T^{tt}$ in the FRW scenario.}
	\label{fig.TttFRW}
	\end{center}
\end{figure}


\begin{figure}[t]
	\begin{center}
	\includegraphics[width=\columnwidth]{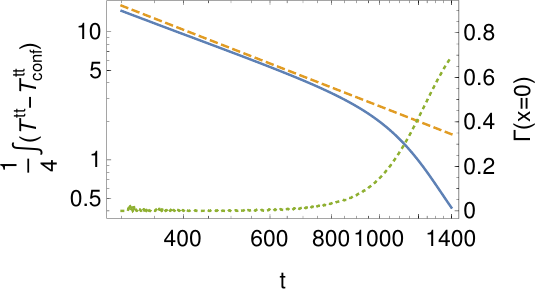}
	\caption{Decay of the averaged $T^{tt}$ in the FRW case (blue curve) compared to power-law decay $t^{-\f{3}{2}}$ (dashed orange curve). We also show the value of $\Gamma$ in the middle of the blob (dotted green curve).}
	\label{fig.TttGamma}
	\end{center}
\end{figure}


\begin{figure}[t]
	\begin{center}
	\hspace{-1cm}\includegraphics[width=0.9\columnwidth]{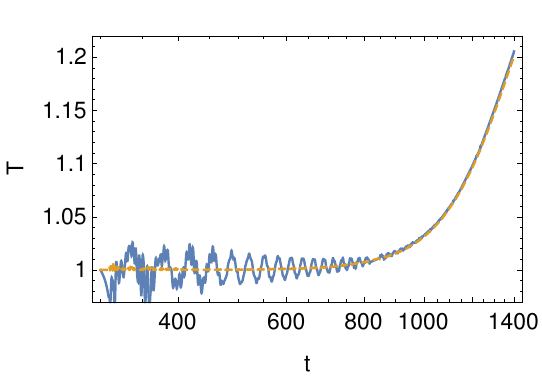}
	\caption{Evolution of the temperature in the middle of the dissolving blob in the FRW scenario (blue curve) compared to the prediction from the virial theorem (dashed orange curve). }
	\label{fig.virialFRW}
	\end{center}
\end{figure}

\medskip

\noindent {\bf Expanding FRW universe.}
As we saw in the preceding discussion, the appearance and subsequent evolution of the hot remnants could be understood on very general grounds as a consequence of dynamics driven by longitudinal (boost-invariant) expansion. Therefore, it is particularly interesting to consider the system in the context of cosmological expansion, in order to check if similar phenomena also appear despite the differences in the character of the expansion. Moreover, a phase transition in the early universe may be the most physically important application of the phenomena discussed in this letter.

Concretely, we analyze our model in an expanding FRW universe
\eq
\label{e.frw}
ds^2 = -dt^2 + a^2(t) d\vec{x}^2 
\eqx
For definiteness, we consider here the radiation dominated phase with $a(t) \sim t^{\f{1}{2}}$, although we checked that similar results hold also e.g. for $a(t) \sim e^{Ht}$ with small~$H$.
Here we work with the physical four dimensional spacetime, 
while for numerical reasons all our domains have planar symmetry. 
Note that the back-reaction of the matter system on the cosmological expansion is not taken into account (c.f.~\cite{Ecker_2023}). 

The increasing scale factor $a(t)$ drives the cooling of the system, so a uniform ``high energy'' phase (with $p(T)=T^4$) in the spacetime~(\ref{e.frw}) expands and cools as $T(t) \sim 1/a(t)$, which keeps the entropy conserved. The plasma becomes supercooled, nucleates bubbles of confined phase, which we mimic by adding Gaussian seeds and just as before leads to the appearance of hot remnants surrounded by the confined phase.

As shown in Fig.~\ref{fig.TttFRW}, we observe the same qualitative stages of the subsequent evolution, first shrinking with the temperature being constant at $T_c$ and then subsequent dissolution.
In the FRW spacetime, one can estimate the shrinking 
of a planar blob 
with
\eq
\partial_t \left( a^3 E\right) +3 a^2 \dot{a} P = 0
\eqx
instead of~(\ref{e.integrated}). This gives
\eq
\label{e.LtFRW}
L(t) \sim \f{1}{a(t)^3}
\eqx
In case of general, nonplanar blobs we would get $\mathrm{Volume}(t) \sim \f{1}{a(t)^3}$ which again conserves entropy (note that here  the volume is measured in the $\vec{x}$ coordinates).
In Fig.~\ref{fig.TttGamma}, we see a very good agreement with~(\ref{e.LtFRW}) in the shrinking phase (i.e. when $\Gm\sim 0$ in the blob).

In order to analyze the dissolution of the remnant, we have to examine the $\gm$ equation of motion
\eq
-\partial_t^2 \gm -\f{3 \dot{a}}{a} \partial_t \gm +\f{1}{a^2} \Delta_x \gm = \f{1}{c} \f{\partial V_\mathrm{TOT}(\gm, T)}{\partial \gm}
\eqx
Neglecting the explicit temporal derivatives, the spatial part can be translated into a virial equation, but modulated by the scale factor $a(t)$. This indicates that the width of the dissolving blob shrinks with $a(t)$ in contrast to staying constant in the boost-invariant case. The virial theorem also leads to the conclusion of a rise of the temperature of the dissolving fluid. This is in very good agreement with the numerical simulations as shown in Fig.~\ref{fig.virialFRW}.
Note that the observed behavior is driven by the expansion of the universe, and qualitatively  different from heated droplets in earlier studies (see~\cite{Cutting:2019zws,Cutting:2022zgd} and references therein) in particular due to the rise of the temperature above $T_c$.


\bigskip

\noindent {\bf Discussion.} In this letter, we explore some generic features of the evolution of holographic Witten model with a confinement-deconfinement phase transition in an expanding setup using an effective QFT boundary description.
We find the appearance of hot remnants of the deconfined phase, which get reheated to the phase transition temperature and stay at that temperature while shrinking in size. Subsequently, we observe a distinct stage of dissolution, where the fraction of the deconfined phase decreases, while starting to heat up. These stages and their properties are quite robust, as they appear both in a boost-invariant expansion in flat space and in cosmological expansion in a FRW universe.

We note that despite the rather complicated and numerically difficult simulations, the appearance and the qualitative features of the hot remnants can be understood quite explicitly using simple physical arguments. We hope that this will facilitate the investigation of their potential role and impact in various applications, in particular phase transitions in the early universe (see~\cite{Hindmarsh:2020hop} and references therein), which seem to be the most promising context for the physics described in the present letter.

In this respect, there are a number of physically relevant possible extensions of the present work, like going to spherical bubble topology in the FRW 
case, as well as the case of very fast expansion.

Apart from direct physical applications, this work also opens up very intriguing questions on the holographic side. The simplified boundary description for the holographic Witten model used in this letter, can be understood as a proxy for the (slowly-varying) regime of Einstein's equations in the bulk. 
From this perspective, the final dissolving phase of the remnant is especially intriguing from the bulk gravitational side, as the dual description entails the disappearance of a black hole and the very final asymptotic stage could be interpreted as a system of glueballs as being a small perturbation on top of the confining geometry.
The precise elucidation of what happens on the bulk gravity side remains as a very interesting open problem.


\bigskip
\noindent{\bf Acknowledgements.} 
We would like to thank Ofer Aharony for discussions and comments on the manuscript.
RJ was supported by a  Priority Research Area DigiWorld grant under the Strategic Programme Excellence Initiative at the Jagiellonian University (Kraków, Poland).
MJ has been supported by an appointment to the JRG Program at the APCTP through the Science and Technology Promotion Fund and Lottery Fund of the Korean Government and by the Korean Local Governments -- Gyeong\-sang\-buk-do Province and Pohang City -- and by the National Research Foundation of Korea (NRF) funded by the Korean government (MSIT) (grant no. 2021R1A2C1010834).
The work of JS was supported in part by a grant 01034816 titled ``String theory reloaded- from fundamental questions to applications'' of the ``Planning and budgeting committee''.


\bibliography{references}

\begin{thebibliography}{30}%
\makeatletter
\providecommand \@ifxundefined [1]{%
 \@ifx{#1\undefined}
}%
\providecommand \@ifnum [1]{%
 \ifnum #1\expandafter \@firstoftwo
 \else \expandafter \@secondoftwo
 \fi
}%
\providecommand \@ifx [1]{%
 \ifx #1\expandafter \@firstoftwo
 \else \expandafter \@secondoftwo
 \fi
}%
\providecommand \natexlab [1]{#1}%
\providecommand \enquote  [1]{``#1''}%
\providecommand \bibnamefont  [1]{#1}%
\providecommand \bibfnamefont [1]{#1}%
\providecommand \citenamefont [1]{#1}%
\providecommand \href@noop [0]{\@secondoftwo}%
\providecommand \href [0]{\begingroup \@sanitize@url \@href}%
\providecommand \@href[1]{\@@startlink{#1}\@@href}%
\providecommand \@@href[1]{\endgroup#1\@@endlink}%
\providecommand \@sanitize@url [0]{\catcode `\\12\catcode `\$12\catcode
  `\&12\catcode `\#12\catcode `\^12\catcode `\_12\catcode `\%12\relax}%
\providecommand \@@startlink[1]{}%
\providecommand \@@endlink[0]{}%
\providecommand \url  [0]{\begingroup\@sanitize@url \@url }%
\providecommand \@url [1]{\endgroup\@href {#1}{\urlprefix }}%
\providecommand \urlprefix  [0]{URL }%
\providecommand \Eprint [0]{\href }%
\providecommand \doibase [0]{https://doi.org/}%
\providecommand \selectlanguage [0]{\@gobble}%
\providecommand \bibinfo  [0]{\@secondoftwo}%
\providecommand \bibfield  [0]{\@secondoftwo}%
\providecommand \translation [1]{[#1]}%
\providecommand \BibitemOpen [0]{}%
\providecommand \bibitemStop [0]{}%
\providecommand \bibitemNoStop [0]{.\EOS\space}%
\providecommand \EOS [0]{\spacefactor3000\relax}%
\providecommand \BibitemShut  [1]{\csname bibitem#1\endcsname}%
\let\auto@bib@innerbib\@empty
\bibitem [{\citenamefont {Janik}\ \emph {et~al.}(2021)\citenamefont {Janik},
  \citenamefont {J{\"a}rvinen},\ and\ \citenamefont
  {Sonnenschein}}]{Janik:2021jbq}%
  \BibitemOpen
  \bibfield  {author} {\bibinfo {author} {\bibfnamefont {R.~A.}\ \bibnamefont
  {Janik}}, \bibinfo {author} {\bibfnamefont {M.}~\bibnamefont
  {J{\"a}rvinen}},\ and\ \bibinfo {author} {\bibfnamefont {J.}~\bibnamefont
  {Sonnenschein}},\ }\bibfield  {title} {\bibinfo {title} {{A simple
  description of holographic domain walls in confining theories \textemdash{}
  extended hydrodynamics}},\ }\href {https://doi.org/10.1007/JHEP09(2021)129}
  {\bibfield  {journal} {\bibinfo  {journal} {JHEP}\ }\textbf {\bibinfo
  {volume} {09}},\ \bibinfo {pages} {129}},\ \Eprint
  {https://arxiv.org/abs/2106.02642} {arXiv:2106.02642 [hep-th]} \BibitemShut
  {NoStop}%
\bibitem [{\citenamefont {Witten}(1998)}]{Witten:1998zw}%
  \BibitemOpen
  \bibfield  {author} {\bibinfo {author} {\bibfnamefont {E.}~\bibnamefont
  {Witten}},\ }\bibfield  {title} {\bibinfo {title} {{Anti-de Sitter space,
  thermal phase transition, and confinement in gauge theories}},\ }\href
  {https://doi.org/10.4310/ATMP.1998.v2.n3.a3} {\bibfield  {journal} {\bibinfo
  {journal} {Adv. Theor. Math. Phys.}\ }\textbf {\bibinfo {volume} {2}},\
  \bibinfo {pages} {505} (\bibinfo {year} {1998})},\ \Eprint
  {https://arxiv.org/abs/hep-th/9803131} {arXiv:hep-th/9803131} \BibitemShut
  {NoStop}%
\bibitem [{\citenamefont {Aharony}\ \emph {et~al.}(2006)\citenamefont
  {Aharony}, \citenamefont {Minwalla},\ and\ \citenamefont
  {Wiseman}}]{Aharony:2005bm}%
  \BibitemOpen
  \bibfield  {author} {\bibinfo {author} {\bibfnamefont {O.}~\bibnamefont
  {Aharony}}, \bibinfo {author} {\bibfnamefont {S.}~\bibnamefont {Minwalla}},\
  and\ \bibinfo {author} {\bibfnamefont {T.}~\bibnamefont {Wiseman}},\
  }\bibfield  {title} {\bibinfo {title} {{Plasma-balls in large N gauge
  theories and localized black holes}},\ }\href
  {https://doi.org/10.1088/0264-9381/23/7/001} {\bibfield  {journal} {\bibinfo
  {journal} {Class. Quant. Grav.}\ }\textbf {\bibinfo {volume} {23}},\ \bibinfo
  {pages} {2171} (\bibinfo {year} {2006})},\ \Eprint
  {https://arxiv.org/abs/hep-th/0507219} {arXiv:hep-th/0507219} \BibitemShut
  {NoStop}%
\bibitem [{\citenamefont {Aharony}\ \emph {et~al.}(2007)\citenamefont
  {Aharony}, \citenamefont {Sonnenschein},\ and\ \citenamefont
  {Yankielowicz}}]{Aharony:2006da}%
  \BibitemOpen
  \bibfield  {author} {\bibinfo {author} {\bibfnamefont {O.}~\bibnamefont
  {Aharony}}, \bibinfo {author} {\bibfnamefont {J.}~\bibnamefont
  {Sonnenschein}},\ and\ \bibinfo {author} {\bibfnamefont {S.}~\bibnamefont
  {Yankielowicz}},\ }\bibfield  {title} {\bibinfo {title} {{A Holographic model
  of deconfinement and chiral symmetry restoration}},\ }\href
  {https://doi.org/10.1016/j.aop.2006.11.002} {\bibfield  {journal} {\bibinfo
  {journal} {Annals Phys.}\ }\textbf {\bibinfo {volume} {322}},\ \bibinfo
  {pages} {1420} (\bibinfo {year} {2007})},\ \Eprint
  {https://arxiv.org/abs/hep-th/0604161} {arXiv:hep-th/0604161} \BibitemShut
  {NoStop}%
\bibitem [{\citenamefont {Gursoy}\ \emph {et~al.}(2008)\citenamefont {Gursoy},
  \citenamefont {Kiritsis}, \citenamefont {Mazzanti},\ and\ \citenamefont
  {Nitti}}]{Gursoy:2008bu}%
  \BibitemOpen
  \bibfield  {author} {\bibinfo {author} {\bibfnamefont {U.}~\bibnamefont
  {Gursoy}}, \bibinfo {author} {\bibfnamefont {E.}~\bibnamefont {Kiritsis}},
  \bibinfo {author} {\bibfnamefont {L.}~\bibnamefont {Mazzanti}},\ and\
  \bibinfo {author} {\bibfnamefont {F.}~\bibnamefont {Nitti}},\ }\bibfield
  {title} {\bibinfo {title} {{Deconfinement and Gluon Plasma Dynamics in
  Improved Holographic QCD}},\ }\href
  {https://doi.org/10.1103/PhysRevLett.101.181601} {\bibfield  {journal}
  {\bibinfo  {journal} {Phys. Rev. Lett.}\ }\textbf {\bibinfo {volume} {101}},\
  \bibinfo {pages} {181601} (\bibinfo {year} {2008})},\ \Eprint
  {https://arxiv.org/abs/0804.0899} {arXiv:0804.0899 [hep-th]} \BibitemShut
  {NoStop}%
\bibitem [{\citenamefont {Gursoy}\ \emph {et~al.}(2009)\citenamefont {Gursoy},
  \citenamefont {Kiritsis}, \citenamefont {Mazzanti},\ and\ \citenamefont
  {Nitti}}]{Gursoy:2008za}%
  \BibitemOpen
  \bibfield  {author} {\bibinfo {author} {\bibfnamefont {U.}~\bibnamefont
  {Gursoy}}, \bibinfo {author} {\bibfnamefont {E.}~\bibnamefont {Kiritsis}},
  \bibinfo {author} {\bibfnamefont {L.}~\bibnamefont {Mazzanti}},\ and\
  \bibinfo {author} {\bibfnamefont {F.}~\bibnamefont {Nitti}},\ }\bibfield
  {title} {\bibinfo {title} {{Holography and Thermodynamics of 5D
  Dilaton-gravity}},\ }\href {https://doi.org/10.1088/1126-6708/2009/05/033}
  {\bibfield  {journal} {\bibinfo  {journal} {JHEP}\ }\textbf {\bibinfo
  {volume} {05}},\ \bibinfo {pages} {033}},\ \Eprint
  {https://arxiv.org/abs/0812.0792} {arXiv:0812.0792 [hep-th]} \BibitemShut
  {NoStop}%
\bibitem [{\citenamefont {Alho}\ \emph {et~al.}(2015)\citenamefont {Alho},
  \citenamefont {Jarvinen}, \citenamefont {Kajantie}, \citenamefont
  {Kiritsis},\ and\ \citenamefont {Tuominen}}]{Alho:2015zua}%
  \BibitemOpen
  \bibfield  {author} {\bibinfo {author} {\bibfnamefont {T.}~\bibnamefont
  {Alho}}, \bibinfo {author} {\bibfnamefont {M.}~\bibnamefont {Jarvinen}},
  \bibinfo {author} {\bibfnamefont {K.}~\bibnamefont {Kajantie}}, \bibinfo
  {author} {\bibfnamefont {E.}~\bibnamefont {Kiritsis}},\ and\ \bibinfo
  {author} {\bibfnamefont {K.}~\bibnamefont {Tuominen}},\ }\bibfield  {title}
  {\bibinfo {title} {{Quantum and stringy corrections to the equation of state
  of holographic QCD matter and the nature of the chiral transition}},\ }\href
  {https://doi.org/10.1103/PhysRevD.91.055017} {\bibfield  {journal} {\bibinfo
  {journal} {Phys. Rev. D}\ }\textbf {\bibinfo {volume} {91}},\ \bibinfo
  {pages} {055017} (\bibinfo {year} {2015})},\ \Eprint
  {https://arxiv.org/abs/1501.06379} {arXiv:1501.06379 [hep-ph]} \BibitemShut
  {NoStop}%
\bibitem [{\citenamefont {Aref'eva}\ and\ \citenamefont
  {Rannu}(2018)}]{Arefeva:2018hyo}%
  \BibitemOpen
  \bibfield  {author} {\bibinfo {author} {\bibfnamefont {I.}~\bibnamefont
  {Aref'eva}}\ and\ \bibinfo {author} {\bibfnamefont {K.}~\bibnamefont
  {Rannu}},\ }\bibfield  {title} {\bibinfo {title} {{Holographic Anisotropic
  Background with Confinement-Deconfinement Phase Transition}},\ }\href
  {https://doi.org/10.1007/JHEP05(2018)206} {\bibfield  {journal} {\bibinfo
  {journal} {JHEP}\ }\textbf {\bibinfo {volume} {05}},\ \bibinfo {pages}
  {206}},\ \Eprint {https://arxiv.org/abs/1802.05652} {arXiv:1802.05652
  [hep-th]} \BibitemShut {NoStop}%
\bibitem [{\citenamefont {Bea}\ \emph {et~al.}(2021)\citenamefont {Bea},
  \citenamefont {Casalderrey-Solana}, \citenamefont {Giannakopoulos},
  \citenamefont {Mateos}, \citenamefont {Sanchez-Garitaonandia},\ and\
  \citenamefont {Zilh\~ao}}]{Bea:2021zsu}%
  \BibitemOpen
  \bibfield  {author} {\bibinfo {author} {\bibfnamefont {Y.}~\bibnamefont
  {Bea}}, \bibinfo {author} {\bibfnamefont {J.}~\bibnamefont
  {Casalderrey-Solana}}, \bibinfo {author} {\bibfnamefont {T.}~\bibnamefont
  {Giannakopoulos}}, \bibinfo {author} {\bibfnamefont {D.}~\bibnamefont
  {Mateos}}, \bibinfo {author} {\bibfnamefont {M.}~\bibnamefont
  {Sanchez-Garitaonandia}},\ and\ \bibinfo {author} {\bibfnamefont
  {M.}~\bibnamefont {Zilh\~ao}},\ }\bibfield  {title} {\bibinfo {title}
  {{Bubble wall velocity from holography}},\ }\href
  {https://doi.org/10.1103/PhysRevD.104.L121903} {\bibfield  {journal}
  {\bibinfo  {journal} {Phys. Rev. D}\ }\textbf {\bibinfo {volume} {104}},\
  \bibinfo {pages} {L121903} (\bibinfo {year} {2021})},\ \Eprint
  {https://arxiv.org/abs/2104.05708} {arXiv:2104.05708 [hep-th]} \BibitemShut
  {NoStop}%
\bibitem [{\citenamefont {Bantilan}\ \emph {et~al.}(2020)\citenamefont
  {Bantilan}, \citenamefont {Figueras},\ and\ \citenamefont
  {Mateos}}]{Bantilan:2020pay}%
  \BibitemOpen
  \bibfield  {author} {\bibinfo {author} {\bibfnamefont {H.}~\bibnamefont
  {Bantilan}}, \bibinfo {author} {\bibfnamefont {P.}~\bibnamefont {Figueras}},\
  and\ \bibinfo {author} {\bibfnamefont {D.}~\bibnamefont {Mateos}},\
  }\bibfield  {title} {\bibinfo {title} {{Real-time Dynamics of Plasma Balls
  from Holography}},\ }\href {https://doi.org/10.1103/PhysRevLett.124.191601}
  {\bibfield  {journal} {\bibinfo  {journal} {Phys. Rev. Lett.}\ }\textbf
  {\bibinfo {volume} {124}},\ \bibinfo {pages} {191601} (\bibinfo {year}
  {2020})},\ \Eprint {https://arxiv.org/abs/2001.05476} {arXiv:2001.05476
  [hep-th]} \BibitemShut {NoStop}%
\bibitem [{\citenamefont {Bigazzi}\ \emph {et~al.}(2020)\citenamefont
  {Bigazzi}, \citenamefont {Caddeo}, \citenamefont {Cotrone},\ and\
  \citenamefont {Paredes}}]{Bigazzi:2020phm}%
  \BibitemOpen
  \bibfield  {author} {\bibinfo {author} {\bibfnamefont {F.}~\bibnamefont
  {Bigazzi}}, \bibinfo {author} {\bibfnamefont {A.}~\bibnamefont {Caddeo}},
  \bibinfo {author} {\bibfnamefont {A.~L.}\ \bibnamefont {Cotrone}},\ and\
  \bibinfo {author} {\bibfnamefont {A.}~\bibnamefont {Paredes}},\ }\bibfield
  {title} {\bibinfo {title} {{Fate of false vacua in holographic first-order
  phase transitions}},\ }\href {https://doi.org/10.1007/JHEP12(2020)200}
  {\bibfield  {journal} {\bibinfo  {journal} {JHEP}\ }\textbf {\bibinfo
  {volume} {12}},\ \bibinfo {pages} {200}},\ \Eprint
  {https://arxiv.org/abs/2008.02579} {arXiv:2008.02579 [hep-th]} \BibitemShut
  {NoStop}%
\bibitem [{\citenamefont {Bigazzi}\ \emph {et~al.}(2021)\citenamefont
  {Bigazzi}, \citenamefont {Caddeo}, \citenamefont {Cotrone},\ and\
  \citenamefont {Paredes}}]{Bigazzi:2020avc}%
  \BibitemOpen
  \bibfield  {author} {\bibinfo {author} {\bibfnamefont {F.}~\bibnamefont
  {Bigazzi}}, \bibinfo {author} {\bibfnamefont {A.}~\bibnamefont {Caddeo}},
  \bibinfo {author} {\bibfnamefont {A.~L.}\ \bibnamefont {Cotrone}},\ and\
  \bibinfo {author} {\bibfnamefont {A.}~\bibnamefont {Paredes}},\ }\bibfield
  {title} {\bibinfo {title} {{Dark Holograms and Gravitational Waves}},\ }\href
  {https://doi.org/10.1007/JHEP04(2021)094} {\bibfield  {journal} {\bibinfo
  {journal} {JHEP}\ }\textbf {\bibinfo {volume} {04}},\ \bibinfo {pages}
  {094}},\ \Eprint {https://arxiv.org/abs/2011.08757} {arXiv:2011.08757
  [hep-ph]} \BibitemShut {NoStop}%
\bibitem [{\citenamefont {Haehl}\ \emph {et~al.}(2015)\citenamefont {Haehl},
  \citenamefont {Loganayagam},\ and\ \citenamefont {Rangamani}}]{Haehl_2015}%
  \BibitemOpen
  \bibfield  {author} {\bibinfo {author} {\bibfnamefont {F.~M.}\ \bibnamefont
  {Haehl}}, \bibinfo {author} {\bibfnamefont {R.}~\bibnamefont {Loganayagam}},\
  and\ \bibinfo {author} {\bibfnamefont {M.}~\bibnamefont {Rangamani}},\
  }\bibfield  {title} {\bibinfo {title} {Adiabatic hydrodynamics: the eightfold
  way to dissipation},\ }\bibfield  {journal} {\bibinfo  {journal} {Journal of
  High Energy Physics}\ }\textbf {\bibinfo {volume} {2015}},\ \href
  {https://doi.org/10.1007/jhep05(2015)060} {10.1007/jhep05(2015)060} (\bibinfo
  {year} {2015})\BibitemShut {NoStop}%
\bibitem [{\citenamefont {Ignatius}\ \emph {et~al.}(1994)\citenamefont
  {Ignatius}, \citenamefont {Kajantie}, \citenamefont {Kurki-Suonio},\ and\
  \citenamefont {Laine}}]{Ignatius:1993qn}%
  \BibitemOpen
  \bibfield  {author} {\bibinfo {author} {\bibfnamefont {J.}~\bibnamefont
  {Ignatius}}, \bibinfo {author} {\bibfnamefont {K.}~\bibnamefont {Kajantie}},
  \bibinfo {author} {\bibfnamefont {H.}~\bibnamefont {Kurki-Suonio}},\ and\
  \bibinfo {author} {\bibfnamefont {M.}~\bibnamefont {Laine}},\ }\bibfield
  {title} {\bibinfo {title} {{The growth of bubbles in cosmological phase
  transitions}},\ }\href {https://doi.org/10.1103/PhysRevD.49.3854} {\bibfield
  {journal} {\bibinfo  {journal} {Phys. Rev. D}\ }\textbf {\bibinfo {volume}
  {49}},\ \bibinfo {pages} {3854} (\bibinfo {year} {1994})},\ \Eprint
  {https://arxiv.org/abs/astro-ph/9309059} {arXiv:astro-ph/9309059}
  \BibitemShut {NoStop}%
\bibitem [{\citenamefont {Hindmarsh}\ \emph {et~al.}(2015)\citenamefont
  {Hindmarsh}, \citenamefont {Huber}, \citenamefont {Rummukainen},\ and\
  \citenamefont {Weir}}]{Hindmarsh:2015qta}%
  \BibitemOpen
  \bibfield  {author} {\bibinfo {author} {\bibfnamefont {M.}~\bibnamefont
  {Hindmarsh}}, \bibinfo {author} {\bibfnamefont {S.~J.}\ \bibnamefont
  {Huber}}, \bibinfo {author} {\bibfnamefont {K.}~\bibnamefont {Rummukainen}},\
  and\ \bibinfo {author} {\bibfnamefont {D.~J.}\ \bibnamefont {Weir}},\
  }\bibfield  {title} {\bibinfo {title} {{Numerical simulations of acoustically
  generated gravitational waves at a first order phase transition}},\ }\href
  {https://doi.org/10.1103/PhysRevD.92.123009} {\bibfield  {journal} {\bibinfo
  {journal} {Phys. Rev. D}\ }\textbf {\bibinfo {volume} {92}},\ \bibinfo
  {pages} {123009} (\bibinfo {year} {2015})},\ \Eprint
  {https://arxiv.org/abs/1504.03291} {arXiv:1504.03291 [astro-ph.CO]}
  \BibitemShut {NoStop}%
\bibitem [{\citenamefont {Janik}\ \emph {et~al.}(2022)\citenamefont {Janik},
  \citenamefont {Jarvinen}, \citenamefont {Soltanpanahi},\ and\ \citenamefont
  {Sonnenschein}}]{Janik:2022wsx}%
  \BibitemOpen
  \bibfield  {author} {\bibinfo {author} {\bibfnamefont {R.~A.}\ \bibnamefont
  {Janik}}, \bibinfo {author} {\bibfnamefont {M.}~\bibnamefont {Jarvinen}},
  \bibinfo {author} {\bibfnamefont {H.}~\bibnamefont {Soltanpanahi}},\ and\
  \bibinfo {author} {\bibfnamefont {J.}~\bibnamefont {Sonnenschein}},\
  }\bibfield  {title} {\bibinfo {title} {{Perfect Fluid Hydrodynamic Picture of
  Domain Wall Velocities at Strong Coupling}},\ }\href
  {https://doi.org/10.1103/PhysRevLett.129.081601} {\bibfield  {journal}
  {\bibinfo  {journal} {Phys. Rev. Lett.}\ }\textbf {\bibinfo {volume} {129}},\
  \bibinfo {pages} {081601} (\bibinfo {year} {2022})},\ \Eprint
  {https://arxiv.org/abs/2205.06274} {arXiv:2205.06274 [hep-th]} \BibitemShut
  {NoStop}%
\bibitem [{\citenamefont {Bjorken}(1983)}]{Bjorken:1982qr}%
  \BibitemOpen
  \bibfield  {author} {\bibinfo {author} {\bibfnamefont {J.~D.}\ \bibnamefont
  {Bjorken}},\ }\bibfield  {title} {\bibinfo {title} {{Highly Relativistic
  Nucleus-Nucleus Collisions: The Central Rapidity Region}},\ }\href
  {https://doi.org/10.1103/PhysRevD.27.140} {\bibfield  {journal} {\bibinfo
  {journal} {Phys. Rev. D}\ }\textbf {\bibinfo {volume} {27}},\ \bibinfo
  {pages} {140} (\bibinfo {year} {1983})}\BibitemShut {NoStop}%
\bibitem [{\citenamefont {Janik}\ and\ \citenamefont
  {Peschanski}(2006)}]{Janik:2005zt}%
  \BibitemOpen
  \bibfield  {author} {\bibinfo {author} {\bibfnamefont {R.~A.}\ \bibnamefont
  {Janik}}\ and\ \bibinfo {author} {\bibfnamefont {R.~B.}\ \bibnamefont
  {Peschanski}},\ }\bibfield  {title} {\bibinfo {title} {{Asymptotic perfect
  fluid dynamics as a consequence of Ads/CFT}},\ }\href
  {https://doi.org/10.1103/PhysRevD.73.045013} {\bibfield  {journal} {\bibinfo
  {journal} {Phys. Rev. D}\ }\textbf {\bibinfo {volume} {73}},\ \bibinfo
  {pages} {045013} (\bibinfo {year} {2006})},\ \Eprint
  {https://arxiv.org/abs/hep-th/0512162} {arXiv:hep-th/0512162} \BibitemShut
  {NoStop}%
\bibitem [{\citenamefont {Chesler}\ and\ \citenamefont
  {Yaffe}(2010)}]{Chesler:2009cy}%
  \BibitemOpen
  \bibfield  {author} {\bibinfo {author} {\bibfnamefont {P.~M.}\ \bibnamefont
  {Chesler}}\ and\ \bibinfo {author} {\bibfnamefont {L.~G.}\ \bibnamefont
  {Yaffe}},\ }\bibfield  {title} {\bibinfo {title} {{Boost invariant flow,
  black hole formation, and far-from-equilibrium dynamics in N = 4
  supersymmetric Yang-Mills theory}},\ }\href
  {https://doi.org/10.1103/PhysRevD.82.026006} {\bibfield  {journal} {\bibinfo
  {journal} {Phys. Rev. D}\ }\textbf {\bibinfo {volume} {82}},\ \bibinfo
  {pages} {026006} (\bibinfo {year} {2010})},\ \Eprint
  {https://arxiv.org/abs/0906.4426} {arXiv:0906.4426 [hep-th]} \BibitemShut
  {NoStop}%
\bibitem [{\citenamefont {Heller}\ \emph {et~al.}(2012)\citenamefont {Heller},
  \citenamefont {Janik},\ and\ \citenamefont {Witaszczyk}}]{Heller:2011ju}%
  \BibitemOpen
  \bibfield  {author} {\bibinfo {author} {\bibfnamefont {M.~P.}\ \bibnamefont
  {Heller}}, \bibinfo {author} {\bibfnamefont {R.~A.}\ \bibnamefont {Janik}},\
  and\ \bibinfo {author} {\bibfnamefont {P.}~\bibnamefont {Witaszczyk}},\
  }\bibfield  {title} {\bibinfo {title} {{The characteristics of thermalization
  of boost-invariant plasma from holography}},\ }\href
  {https://doi.org/10.1103/PhysRevLett.108.201602} {\bibfield  {journal}
  {\bibinfo  {journal} {Phys. Rev. Lett.}\ }\textbf {\bibinfo {volume} {108}},\
  \bibinfo {pages} {201602} (\bibinfo {year} {2012})},\ \Eprint
  {https://arxiv.org/abs/1103.3452} {arXiv:1103.3452 [hep-th]} \BibitemShut
  {NoStop}%
\bibitem [{\citenamefont {Heller}\ \emph {et~al.}(2013)\citenamefont {Heller},
  \citenamefont {Janik},\ and\ \citenamefont {Witaszczyk}}]{Heller:2013fn}%
  \BibitemOpen
  \bibfield  {author} {\bibinfo {author} {\bibfnamefont {M.~P.}\ \bibnamefont
  {Heller}}, \bibinfo {author} {\bibfnamefont {R.~A.}\ \bibnamefont {Janik}},\
  and\ \bibinfo {author} {\bibfnamefont {P.}~\bibnamefont {Witaszczyk}},\
  }\bibfield  {title} {\bibinfo {title} {{Hydrodynamic Gradient Expansion in
  Gauge Theory Plasmas}},\ }\href
  {https://doi.org/10.1103/PhysRevLett.110.211602} {\bibfield  {journal}
  {\bibinfo  {journal} {Phys. Rev. Lett.}\ }\textbf {\bibinfo {volume} {110}},\
  \bibinfo {pages} {211602} (\bibinfo {year} {2013})},\ \Eprint
  {https://arxiv.org/abs/1302.0697} {arXiv:1302.0697 [hep-th]} \BibitemShut
  {NoStop}%
\bibitem [{\citenamefont {Gursoy}\ \emph {et~al.}(2016)\citenamefont {Gursoy},
  \citenamefont {Jarvinen},\ and\ \citenamefont {Policastro}}]{Gursoy:2015nza}%
  \BibitemOpen
  \bibfield  {author} {\bibinfo {author} {\bibfnamefont {U.}~\bibnamefont
  {Gursoy}}, \bibinfo {author} {\bibfnamefont {M.}~\bibnamefont {Jarvinen}},\
  and\ \bibinfo {author} {\bibfnamefont {G.}~\bibnamefont {Policastro}},\
  }\bibfield  {title} {\bibinfo {title} {{Late time behavior of non-conformal
  plasmas}},\ }\href {https://doi.org/10.1007/JHEP01(2016)134} {\bibfield
  {journal} {\bibinfo  {journal} {JHEP}\ }\textbf {\bibinfo {volume} {01}},\
  \bibinfo {pages} {134}},\ \Eprint {https://arxiv.org/abs/1507.08628}
  {arXiv:1507.08628 [hep-th]} \BibitemShut {NoStop}%
\bibitem [{\citenamefont {Aragon\`es~Fontbot\'e}\ \emph
  {et~al.}(2024)\citenamefont {Aragon\`es~Fontbot\'e}, \citenamefont {Mateos},
  \citenamefont {Mart\'\i{}n}, \citenamefont {van~der Schee},\ and\
  \citenamefont {Subils}}]{AragonesFontbote:2024lor}%
  \BibitemOpen
  \bibfield  {author} {\bibinfo {author} {\bibfnamefont {M.}~\bibnamefont
  {Aragon\`es~Fontbot\'e}}, \bibinfo {author} {\bibfnamefont {D.}~\bibnamefont
  {Mateos}}, \bibinfo {author} {\bibfnamefont {G.~P.}\ \bibnamefont
  {Mart\'\i{}n}}, \bibinfo {author} {\bibfnamefont {W.}~\bibnamefont {van~der
  Schee}},\ and\ \bibinfo {author} {\bibfnamefont {J.~G.}\ \bibnamefont
  {Subils}},\ }\bibfield  {title} {\bibinfo {title} {{Cosmic censorship in a
  (dual) collider}},\ }\href@noop {} {\  (\bibinfo {year} {2024})},\ \Eprint
  {https://arxiv.org/abs/2411.17806} {arXiv:2411.17806 [hep-th]} \BibitemShut
  {NoStop}%
\bibitem [{\citenamefont {Li}\ \emph {et~al.}(2023)\citenamefont {Li},
  \citenamefont {Wang},\ and\ \citenamefont {Yuwen}}]{Li:2023xto}%
  \BibitemOpen
  \bibfield  {author} {\bibinfo {author} {\bibfnamefont {L.}~\bibnamefont
  {Li}}, \bibinfo {author} {\bibfnamefont {S.-J.}\ \bibnamefont {Wang}},\ and\
  \bibinfo {author} {\bibfnamefont {Z.-Y.}\ \bibnamefont {Yuwen}},\ }\bibfield
  {title} {\bibinfo {title} {{Bubble expansion at strong coupling}},\ }\href
  {https://doi.org/10.1103/PhysRevD.108.096033} {\bibfield  {journal} {\bibinfo
   {journal} {Phys. Rev. D}\ }\textbf {\bibinfo {volume} {108}},\ \bibinfo
  {pages} {096033} (\bibinfo {year} {2023})},\ \Eprint
  {https://arxiv.org/abs/2302.10042} {arXiv:2302.10042 [hep-th]} \BibitemShut
  {NoStop}%
\bibitem [{\citenamefont {Wang}\ \emph {et~al.}(2024)\citenamefont {Wang},
  \citenamefont {Yuwen}, \citenamefont {Hao},\ and\ \citenamefont
  {Wang}}]{Wang:2023lam}%
  \BibitemOpen
  \bibfield  {author} {\bibinfo {author} {\bibfnamefont {J.-C.}\ \bibnamefont
  {Wang}}, \bibinfo {author} {\bibfnamefont {Z.-Y.}\ \bibnamefont {Yuwen}},
  \bibinfo {author} {\bibfnamefont {Y.-S.}\ \bibnamefont {Hao}},\ and\ \bibinfo
  {author} {\bibfnamefont {S.-J.}\ \bibnamefont {Wang}},\ }\bibfield  {title}
  {\bibinfo {title} {{General bubble expansion at strong coupling}},\ }\href
  {https://doi.org/10.1103/PhysRevD.109.096012} {\bibfield  {journal} {\bibinfo
   {journal} {Phys. Rev. D}\ }\textbf {\bibinfo {volume} {109}},\ \bibinfo
  {pages} {096012} (\bibinfo {year} {2024})},\ \Eprint
  {https://arxiv.org/abs/2311.07347} {arXiv:2311.07347 [hep-ph]} \BibitemShut
  {NoStop}%
\bibitem [{\citenamefont {Clough}(2021)}]{Clough:2021qlv}%
  \BibitemOpen
  \bibfield  {author} {\bibinfo {author} {\bibfnamefont {K.}~\bibnamefont
  {Clough}},\ }\bibfield  {title} {\bibinfo {title} {{Continuity equations for
  general matter: applications in numerical relativity}},\ }\href
  {https://doi.org/10.1088/1361-6382/ac10ee} {\bibfield  {journal} {\bibinfo
  {journal} {Class. Quant. Grav.}\ }\textbf {\bibinfo {volume} {38}},\ \bibinfo
  {pages} {167001} (\bibinfo {year} {2021})},\ \Eprint
  {https://arxiv.org/abs/2104.13420} {arXiv:2104.13420 [gr-qc]} \BibitemShut
  {NoStop}%
\bibitem [{\citenamefont {Ecker}\ \emph {et~al.}(2023)\citenamefont {Ecker},
  \citenamefont {Kiritsis},\ and\ \citenamefont {van~der Schee}}]{Ecker_2023}%
  \BibitemOpen
  \bibfield  {author} {\bibinfo {author} {\bibfnamefont {C.}~\bibnamefont
  {Ecker}}, \bibinfo {author} {\bibfnamefont {E.}~\bibnamefont {Kiritsis}},\
  and\ \bibinfo {author} {\bibfnamefont {W.}~\bibnamefont {van~der Schee}},\
  }\bibfield  {title} {\bibinfo {title} {Dynamical inflaton coupled to strongly
  interacting matter},\ }\bibfield  {journal} {\bibinfo  {journal} {Physical
  Review Letters}\ }\textbf {\bibinfo {volume} {130}},\ \href
  {https://doi.org/10.1103/physrevlett.130.251001}
  {10.1103/physrevlett.130.251001} (\bibinfo {year} {2023})\BibitemShut
  {NoStop}%
\bibitem [{\citenamefont {Cutting}\ \emph {et~al.}(2020)\citenamefont
  {Cutting}, \citenamefont {Hindmarsh},\ and\ \citenamefont
  {Weir}}]{Cutting:2019zws}%
  \BibitemOpen
  \bibfield  {author} {\bibinfo {author} {\bibfnamefont {D.}~\bibnamefont
  {Cutting}}, \bibinfo {author} {\bibfnamefont {M.}~\bibnamefont {Hindmarsh}},\
  and\ \bibinfo {author} {\bibfnamefont {D.~J.}\ \bibnamefont {Weir}},\
  }\bibfield  {title} {\bibinfo {title} {{Vorticity, kinetic energy, and
  suppressed gravitational wave production in strong first order phase
  transitions}},\ }\href {https://doi.org/10.1103/PhysRevLett.125.021302}
  {\bibfield  {journal} {\bibinfo  {journal} {Phys. Rev. Lett.}\ }\textbf
  {\bibinfo {volume} {125}},\ \bibinfo {pages} {021302} (\bibinfo {year}
  {2020})},\ \Eprint {https://arxiv.org/abs/1906.00480} {arXiv:1906.00480
  [hep-ph]} \BibitemShut {NoStop}%
\bibitem [{\citenamefont {Cutting}\ \emph {et~al.}(2022)\citenamefont
  {Cutting}, \citenamefont {Vilhonen},\ and\ \citenamefont
  {Weir}}]{Cutting:2022zgd}%
  \BibitemOpen
  \bibfield  {author} {\bibinfo {author} {\bibfnamefont {D.}~\bibnamefont
  {Cutting}}, \bibinfo {author} {\bibfnamefont {E.}~\bibnamefont {Vilhonen}},\
  and\ \bibinfo {author} {\bibfnamefont {D.~J.}\ \bibnamefont {Weir}},\
  }\bibfield  {title} {\bibinfo {title} {{Droplet collapse during strongly
  supercooled transitions}},\ }\href
  {https://doi.org/10.1103/PhysRevD.106.103524} {\bibfield  {journal} {\bibinfo
   {journal} {Phys. Rev. D}\ }\textbf {\bibinfo {volume} {106}},\ \bibinfo
  {pages} {103524} (\bibinfo {year} {2022})},\ \Eprint
  {https://arxiv.org/abs/2204.03396} {arXiv:2204.03396 [astro-ph.CO]}
  \BibitemShut {NoStop}%
\bibitem [{\citenamefont {Hindmarsh}\ \emph {et~al.}(2021)\citenamefont
  {Hindmarsh}, \citenamefont {L\"uben}, \citenamefont {Lumma},\ and\
  \citenamefont {Pauly}}]{Hindmarsh:2020hop}%
  \BibitemOpen
  \bibfield  {author} {\bibinfo {author} {\bibfnamefont {M.~B.}\ \bibnamefont
  {Hindmarsh}}, \bibinfo {author} {\bibfnamefont {M.}~\bibnamefont {L\"uben}},
  \bibinfo {author} {\bibfnamefont {J.}~\bibnamefont {Lumma}},\ and\ \bibinfo
  {author} {\bibfnamefont {M.}~\bibnamefont {Pauly}},\ }\bibfield  {title}
  {\bibinfo {title} {{Phase transitions in the early universe}},\ }\href
  {https://doi.org/10.21468/SciPostPhysLectNotes.24} {\bibfield  {journal}
  {\bibinfo  {journal} {SciPost Phys. Lect. Notes}\ }\textbf {\bibinfo {volume}
  {24}},\ \bibinfo {pages} {1} (\bibinfo {year} {2021})},\ \Eprint
  {https://arxiv.org/abs/2008.09136} {arXiv:2008.09136 [astro-ph.CO]}
  \BibitemShut {NoStop}%
\end{thebibliography}%

\clearpage

\appendix*
\setcounter{equation}{0}

\section*{Supplemental material}

\subsection{Details on the simplified model}

We discuss here some details of the extended hydrodynamic model of equation~(\ref{e.two}) in the letter (which we also refer to as the simplified model). This model was introduced in~\cite{Janik:2021jbq} motivated by the numerical results in the Witten's model obtained in~\cite{Aharony:2005bm}, i.e., the solution for the static domain wall between the confined and deconfined phases (the AMW solution). More precisely, the terms in the action of the hydrodynamic model were motivated by fits to the energy-momentum tensor extracted from the AMW solution, and the final hydrodynamic model was seen to reproduce the data for the energy-momentum tensor to an extremely high precision given the simplicity of the model.

In the letter we used both a 2+1 dimensional formalism (to describe expanding quark-gluon plasma) and a 3+1 dimensional formalism (to describe a FRW universe). We first discuss the model in 2+1 dimensions.

In the Lagrangian~(\ref{e.two})  we have chosen the units such that the critical temperature of the deconfinement transition $T_c=1$, and the pressure, which is that of a conformal theory, is given by $p(T)=T^4$.
 
The constants $c$ and $q_*$, which control the value of the surface tension and the width of the domain wall respectively, were fitted to the AMW solution. This fit gives $q_*/(2\pi T_c) \approx 0.682$, but because the value of $q_*$ can be absorbed into rescalings of the space-time coordinates, we can set $q_*=1$ without loss of generality. For this choice, the AMW value for $c$ is $c \approx 5.892$.

The expressions for the energy-momentum tensors of the confined and deconfined phases in equation~(2) of the letter are given as
\eqn \label{eq:Tconf}
 T_{\mu\nu}^\mathrm{confined} &=& \eta_{\mu\nu} = \mathrm{diag}(-1,1,1)\\
 T_{\mu\nu}^\mathrm{deconfined} &=& p(T)\eta_{\mu\nu} + 4 p(T) u_\mu u_\nu \label{eq:Tdeconf}
\eqnx
where $u_\mu$ is the fluid velocity satisfying $\eta^{\mu\nu}u_\mu u_\nu=-1$ and we eliminated the energy density by using the relation $\eps(T) =3 p(T)$ for conformal theories. 
The functional form of the mixing fraction $\Gm(\gm)$ of the deconfined and confined phases in~\eqref{e.one}, $\Gm(\gm)=\gm^2(3-2\gm)$, was motivated by the profile of the AMW domain wall. 

Here we chose to write the tensor for only for the final 2+1 terms in the system, i.e., we did not write down the components in the third, compactified spatial direction. 
The space-time coordinates are then $(t,x,y)$. We will choose the planar domain wall to lie at fixed value of the coordinate $x$, so that our solutions will be independent of the other spatial coordinate(s). At critical temperature and for fluid at rest the deconfined tensor $T_{\mu\nu}^\mathrm{deconfined} = \eta_{\mu\nu}+4\delta_\mu^t\delta_\nu^t$ in our conventions. For the ``domain wall'' contribution to the energy-momentum tensor we find
\eqn
 T^\Sigma_{\mu\nu}&=& c\, \partial_\mu\gm\partial_\nu\gm \\\nonumber
 &&-\frac{c}{2} \left[(\partial \gm)^2+ q_*^2\,T^{\beta(\gm)}\,\gm^2(1-\gm)^2\right]\eta_{\mu\nu} \\\nonumber
 && -c\, q_*^2\, \left(1+\Gamma(\gm)\right)\,T^{\beta(\gm)}\,\gm^2(1-\gm)^2 u_\mu u_\nu
\eqnx
where 
\eq
\beta(\gm)= 2(1+\Gamma(\gm))
\eqx
is also obtained by comparing to the AMW data~\cite{Janik:2021jbq}.
Note that we assumed for that the kinetic terms are independent of $T$, meaning $\alpha(\gm) = 0$ in the notation of~\cite{Janik:2021jbq}. This is the simplest choice which gives a good description of the domain wall data from the AMW solution.  

Let us then discuss the 3+1 dimensional setup. The AMW solution is also available at one dimension higher~\cite{Aharony:2005bm}, so the simplest option appears to be to simply refit our model to this higher dimensional data. However, this data describes a 4+1 dimensional field theory with one spatial dimension compactified, which leads to the equation of state being $p(T)=T^5$. This equation of state is not favourable for our FRW application. Therefore we opt for a phenomenological approach, where we take the equation of state to be $p(T)=T^4$, use the value $c \approx 5.892$ for the parameter controlling the domain wall tension obtained from the lower dimensional fit, and simply add an extra dimension in~\eqref{eq:Tconf} and~\eqref{eq:Tdeconf} by extending the metric to be $\eta_{\mu\nu} = \mathrm{diag}(-1,1,1,1)$. Moreover, given that the form of $\beta(\gm)$ is no longer strictly determined by any fit to domain wall data, we choose a simpler option $\beta(\gm)=4$ which is motivated by the analysis of the higher dimensional AMW data~\cite{Janik:2021jbq}.

\subsection{Entropy current}

We provide here a construction of a conserved (as we do not incorporate dissipation) entropy current in our model:
\eq
\label{e.model}
\LL = \f{c}{2} (\partial \gm)^2 + V_\mathrm{TOT}(\gm, T)
\eqx
where
\eq
V_\mathrm{TOT}(\gm, T) = -(1-\Gm) p(T) -\Gm + \f{c}{2} q_*^2 T^\beta \gm^2 (1-\gm)^2 +1
\eqx
For constant $\gm$, $\LL$ can be intuitively identified with the free energy $F$ of the system. Then a natural formula for the entropy density follows from the thermodynamic formula
\eq
\label{e.entropygen}
s \equiv - \f{\partial F}{\partial T} = -\partial_T V_\mathrm{TOT}(\gm, T)
\eqx
Defining the entropy current in a standard hydrodynamical way through $j^\mu =s u^\mu$, we have to verify the conservation law $\partial_\mu j^\mu=0$.

The equation of motion of our model are
\eqn
c\, \partial^2 \gm -V_\mathrm{TOT}' &=& 0 \\
\partial_\mu \left( T^{\mu\nu}_\mathrm{std} + T^{\mu\nu}_\mathrm{extra} \right) &=& 0
\eqnx
where $T^{\mu\nu}_\mathrm{std}$ is the standard energy momentum tensor for the model~(\ref{e.model}) treating the temperature as a conventional external field
\eq
T^{\mu\nu}_\mathrm{std} = c \partial_\mu\gm \partial_\nu\gm - \left( \f{1}{2} c(\partial \gm)^2 +V_\mathrm{TOT} \right) \eta_{\mu\nu}
\eqx
while $T^{\mu\nu}_\mathrm{extra}$ comprises the additional terms proportional to $u^\mu u^\nu$ which arise from the $T$ derivatives of the Lagrangian following the construction of~\cite{Haehl_2015}. We can concisely write them as
\eq
T^{\mu\nu}_\mathrm{extra} = -\left( T \partial_T V_\mathrm{TOT}\right) u^\mu u^\nu
\eqx
Using the $\gm$ equation of motion we have
\eq
\partial_\mu T^{\mu\nu}_\mathrm{std} = - \partial_T V_\mathrm{TOT} \partial^\nu T
\eqx
The total conservation of energy-momentum $\partial_\mu T^{\mu\nu}_\mathrm{std} + \partial_\mu T^{\mu\nu}_\mathrm{extra}=0$ then yields
\eq
-\partial_\mu \left[ \left( T \partial_T V_\mathrm{TOT}\right) u^\mu u^\nu \right] - \partial_T V_\mathrm{TOT} \partial^\nu T = 0
\eqx 
Contracting with $u_\nu$, using $u^2=-1$ gives finally entropy conservation
\eq
\partial_\mu j^\mu = \partial_\mu \left( s\, u^\mu \right) = 0
\eqx
with the entropy density $s=-\partial_T V_\mathrm{TOT}$ as given by~(\ref{e.entropygen}).

\subsection*{Some details on the numerical simulations}

For this letter, we carried out dynamical time-dependent simulations in the simplified model. We discuss here some details of the simulations. We searched for solutions depending on one time coordinate (either the proper time $\tau$ or the time coordinate of the FRW metric) and one spatial coordinate $x$. 

Due to the additional scalar directly using existing advanced hydrodynamic codes is not possible. Therefore we simply reduced the equation of motion to a set of four equations with first-order time derivatives for four fields: the rescaled temperature, $T_s = T \sqrt{1-\gamma}$, the boost parameter $\theta$ of the fluid velocity in $x$-direction, the scalar field $\gamma$, and its time derivative $\partial_t \gamma$. Time evolving this system is in principle straightforward.
However, as it turns out, while evolving the hydrodynamic sector of the model is rather simple, the additional scalar leads to numerical difficulties. In particular, the evolution in the confined regions, where $\gamma \approx 1$, is tricky. To regularize the evolution, we first removed all singularities from our equations of motion by adding small numerical cutoffs. These were the following:
\begin{enumerate}
    \item The point $\gamma=1$. We replaced factors of $1/(1-\gm)$ in the equations of motion by $(1-\gm)/[(1-\gm)^2+\eps_\gm]$.
    \item The point $T_s=0$. We replaced non-integer powers of $T_s$ by regulated expressions, e.g., $T_s^{f(\gm)}$ by $(T_s^2+\eps_T)^{f(\gm)/2}$.
    \item The point $J=0$, where $J$ is the Jacobian arising when solving for the leading time derivatives in the hydrodynamic sector. We replaced factors of $1/J$ by $J/(J^2+\eps_J)$.
\end{enumerate}
Moreover, the map $\Gamma(\gm) = \gm^2(3-2\gm)$ for the mixing fraction actually only makes sense for $0<\gm<1$. Numerical simulations often venture outside this domain. We used a simple extension of this map that does not lead to unphysical solutions: $\Gm=1$ whenever $\gm>1$ and $\Gm=0$ whenever $\gm<0$. Apart from these regulating steps, we were careful to write the final equations in a form that avoids large numerical cancellations between various terms. 

As for the actual numerics, the simulated system was placed in a box with width $L=120$ in the $x$-direction with periodic boundary conditions. As explained above, the units of space and time were fixed such that $q_*=1$, which sets the width of domain walls roughly to one. We introduced a spatial grid with $N=1200$ grid points, so that the spacing between grid points was $\Delta x = 0.1$. Spatial derivatives in the equations of motion were estimated by using sixth order discrete approximations. The time evolution was done by using a fourth-order Runge-Kutta method (RK4) with a step size $\Delta t = 0.125$.

In addition, a relatively strong numerical filter was needed to remove an instability arising from the regions with confined phase. This filter would in essence do a Fourier transform and remove the highest modes in the solution. However due to the complicated spatial dependence of the solutions simply cutting off globally the highest Fourier modes would not work (but would quickly generate noise in the regions where the evolution is eventless). Therefore we applied a local filter, whose width was chosen to be 11 grid points, and which would filter typically the upper half of the Fourier modes. For smoothness, the filter was multiplied by a windowing function and subsequently convoluted with the data after each RK step. That is, the same filter was applied at all values of $x$ and for all four functions.

Even after applying these regulators numerical evolution would often turn out to be unstable. It would be interesting to try to develop a more elegant and stable code for the evolution of the system.

We carried out four different simulations in the setup with expanding plasma. The initial conditions for these simulations were the following:
\begin{enumerate}
    \item Initial condition at $\tau=300$ with $T=0.95 T_c$ and $\gm=0$ (deconfined phase) except for two Gaussian perturbations in $\gm$, centered at $x \approx \pm 10$ with amplitude $0.75$ and width $4.4$. The velocity and the time derivative of $\gm$ were set to zero initially. This simulation was used to generate Fig.~\ref{fig.twogaussian}.
    \item Initial condition for a blob of deconfined matter with $T=0.95 T_c$ and width $60$ at $\tau=300$. That is, $\gm$ was chosen to be nonzero only for $|x|\lesssim 30$, with the blob limited by smooth $\tanh$ steps having widths equal to one. To be precise, we chose the rescaled temperature to be $T_s=0.95 T_c$ in the deconfined phase region and $T_s=0.2 T_c$ in the confined region with $\tanh$ steps in the middle. This value of $T_s$ in the confined phase was chosen because it led to a smooth simulation -- notice that in this phase the temperature has no physical meaning as the hydrodynamic sector is turned off, and the details of the simulation in the blob region are insensitive to this value. This simulation was used to generate the plots in Fig.~\ref{fig.energytemp}. The rather large width was chosen to make the propagating waves inside the blob better visible. 
    \item Initial condition for a blob of deconfined matter with $T=T_c$ and width $40$ at $\tau=50$. Profiles were chosen similarly to the second simulation, but we added also an initial velocity profile (i.e., $\theta$) corresponding to a uniformly shrinking blob. This was necessary to damp the initial oscillations of the blob that would otherwise spoil the simulation. In the absence of initial velocity, these oscillations would be enhanced with respect to the second simulation due to the relatively low initial $\tau$. This simulation was used for Figs.~\ref{fig.size} and~\ref{fig.gamma_dissolve}.
    \item Initial conditions as in the third simulation, but with a narrower width $20$. This width was chosen to see better the effect of the decaying $\gm$ profile. The simulation was used in Fig.~\ref{fig.vanishing}.
\end{enumerate}

We also ran a simulation in the FRW scenario with the following initial condition. We chose $a(t)=\sqrt{t/t_0}$ with $t_0=300$, and set $T=T_c$ at $t=t_0$. The profiles of $\gm$ and $T_s$ were chosen to correspond to a deconfined blob with width $16$ in the same way as in the second simulation above (i.e., also with zero initial velocity). This simulation was used to generate plots in Figs.~\ref{fig.TttFRW},~\ref{fig.TttGamma} and~\ref{fig.virialFRW}.

Finally, we carried out checks of convergence for all numerical simulations. That is, we varied the regulators $\eps_i$, the parameters of the noise filter, grid step sizes, as well as the order of the discrete derivatives, and checked whether the results change. We found that the results in Figs.~\ref{fig.twogaussian},~\ref{fig.energytemp},~\ref{fig.size} and~\ref{fig.gamma_dissolve} have fully converged. The results in the other Figs.~\ref{fig.vanishing},~\ref{fig.TttFRW},~\ref{fig.TttGamma} and~\ref{fig.virialFRW} suffer from a mild uncertainty in the timescale where the $\gm$ field profile starts to dissolve. This timescale turns out to be sensitive to the filter we are using, and some mild dependence remains even at the weakest filter settings where simulations are possible. Note however that even the simulations presented in these figures reproduce well other expected properties such as conservation laws and the predictions of the virial theorem.

\end{document}